\documentstyle[preprint,aps,psfig]{revtex}
\newcommand{\be}{\begin{equation}}
\newcommand{\ee}{\end{equation}}
\newcommand{\bea}{\begin{eqnarray}}
\newcommand{\eea}{\end{eqnarray}}
\tightenlines
\draft 
\preprint{LBL-49562} 
\begin{document}
\title{Azimuthal asymmetry of $J/\psi$ suppression
in non-central heavy-ion collisions}
  
\author{Xin-Nian Wang$^a$ and Feng Yuan$^b$}
\address{$^a$ Nuclear Science Division, MS 70-319,
  Lawrence Berkley National Laboratory, Berkeley, California 94720}
\address{$^b$Institut f\"ur Theoretische Physik der Universit\"at, 
        Philosophenweg 19, D-69120 Heidelberg, Germany}

\maketitle    
%\date{\today}

\begin{abstract}

The azimuthal asymmetry of $J/\psi$ suppression in non-central 
heavy-ion collisions is studied within a dynamic model of $J/\psi$ 
suppression in a deconfined partonic medium. Within this model, $J/\psi$
suppression in heavy-ion collisions is caused mainly by the initial
state nuclear absorption and dissociation via gluon-$J/\psi$ scattering
in deconfined partonic medium. Only the second mechanism gives arise to
azimuthal asymmetry of the final $J/\psi$ production. We demonstrate
that if there is an onset of suppression by quark-gluon plasma (QGP) in 
the NA50 data, it must be accompanied by the non-vanishing azimuthal
asymmetry. Using the same critical density above which the QGP effect 
enters, we predict the azimuthal asymmetric coefficient $v_2$ as well 
as the survival probability for $J/\psi$ at the RHIC energy.

\pacs{PACS number(s): 12.38.Mh; 24.85.+p; 25.75.-q}
\end{abstract}

In the search for quark-gluon plasma (QGP), $J/\psi$ suppression has 
been proposed as one of the promising signals \cite{ms} of the deconfinement
in high-energy heavy-ion collisions. Because of the color screening
effect in a quark-gluon plasma, the linear confining potential in vacuum 
that binds two heavy quarks to form a quarkonium disappears so that
it can be easily broken up causing suppression of the $J/\psi$ production.
The problem in heavy-ion collisions is however complicated by other
competing mechanisms such as initial nuclear absorption \cite{ks}
and hadronic dissociation \cite{review}. While recent precision data 
from the NA50 \cite{na50} experiment at the CERN SPS energies clearly show
anomalous suppression unexplained by the normal initial nuclear absorption,
there are still much debates about the exact nature of the anomalous 
suppression \cite{blaizot,capella}, whether it is caused by the formation
of QGP or dissociation by ordinary hadronic matter.

We propose in this letter the study of azimuthal asymmetry of $J/\psi$
production \cite{hm} as additional measurements to distinguish different 
competing mechanism of $J/\psi$ suppression. Since the initial state
interactions such as nuclear absorption or nuclear shadowing of gluon 
distribution has no preference over the azimuthal direction they will 
not have any contribution to the azimuthal anisotropy of the $J/\psi$
production. Only suppression by the final state interaction with
the produced medium will cause significant azimuthal anisotropy in
the final $J/\psi$ distribution in the transverse direction. If the
centrality dependence of the $J/\psi$ suppression additional to the
initial nuclear absorption is caused by formation of QGP, it must be 
accompanied by a sudden onset of the azimuthal anisotropy. On the
other hand, a hadronic absorption scenario would give a continuous
centrality dependence of the azimuthal anisotropy. In this letter, 
we will study the centrality dependence of both the averaged
$J/\psi$ suppression factor and the azimuthal anisotropy with a model
in which $J/\psi$ suppression is caused by initial nuclear absorption
and final state dissociation by QGP above a critical density. Using
parameters from fitting the NA50 data, we will also give 
predictions for $J/\psi$ suppression and its azimuthal anisotropy
at the RHIC energies.

We follow the microdynamic approach of $J/\psi$ suppression\cite{xw}, 
in which the $J/\psi$ suppression is caused by gluonic dissociation.
The gluonic dissociation cross section in this model depends strongly
on the gluon density and the averaged energy of the gluons in the medium. 
In a hadron gas, one finds \cite{ks} the effective 
hadron-$J/\psi$ cross section to be extremely small because of the small 
gluon distribution inside a thermal hadron. Furthermore, average energy 
carried by the gluon inside a thermal hadron is much smaller than the 
threshold energy of $J/\psi$ dissociation. This further suppresses 
hadron-$J/\psi$ dissociation cross section.
We should emphasize that such an assumption is model dependent. Results of
other theoretical 
calculations \cite{matinian,haglin,linko,wong} of 
hadron-$J/\psi$ dissociation
also vary widely, generally potential models giving large cross
sections while effective interaction models giving much smaller values.
For consistency we will assume a small hadron-$J/\psi$ cross section within
the gluon dissociation model. Therefore, we can neglect
the $J/\psi$ dissociation in a hadron gas which exist in the
peripheral heavy-ion collisions and the late stage of central collisions.

Different from a normal hadron gas, a deconfined partonic system contains
large number of much harder gluons which can easily break up a $J/\psi$.
This will be the only mechanism for final state suppression of $J/\psi$ in
our paper. Such a gluon dissociation model can be considered as a
micro-dynamic model of screening of the potential between heavy quarks
in a quark-gluon plasma \cite{ms,wong,digal}. 
In the screened potential model, a $J/\psi$ is considered destroyed if 
it is produced in a plasma that
is above the critical temperature. On the other hand, the micro-dynamic
model still gives a non-vanishing survival probability that depends
on the temperature of propagation length.

The perturbative calculations predict the gluon-$J/\psi$ dissociation
cross section\cite{peskin,kharzeev},
\begin{equation}
\sigma_\psi(q^0)=N_0 \frac{(q^0/\epsilon_0-1)^{3/2}}{(q^0/\epsilon_0)^5},
\end{equation}
where
$$N_0=\frac{2\pi}{3} (\frac
{32}{3})^2(\frac{16\pi}{3g_s^2})\frac{1}{m_Q^2}.
$$
Here $g_s$ is the strong coupling constant, $m_Q$ is
charm quark mass, and $q^0$ is the gluon energy in the $J/\psi$ rest
frame. To break up a $J/\psi$, $q^0$ must be larger than the
 binding energy $\epsilon_0$. Using similar approach, we have 
calculated the dissociation cross section for $P$-wave states by gluons,
\begin{equation}
\sigma_\chi(q^0)=4 N_0 \frac{(q^0/\epsilon_\chi-1)^{1/2}
(9(q^0/\epsilon_\chi)^2-20(q^0/\epsilon_\chi)+12)}
{(q^0/\epsilon_\chi)^7},
\end{equation}
where $\epsilon_\chi$ is the binding energy of the $P$-wave
state. For $\chi_c$, it is about $0.250GeV$. Because of the small
value of the binding energy, the validity of the perturbative 
calculations for the gluonic dissociation cross section of $\chi_c$ 
might be questionable. So the above formula can only be considered
more phenomenological. However, keeping this point in mind,
we can see that the above expression still qualitatively 
reflects the fact that $\chi_c$ states are easier to be broken 
up than $J/\psi$, because of the much lower energy threshold and 
the overall larger factor of $4$.  Therefore, this perturbative 
calculation gives us a reasonable estimate and
guides us to include its contribution for a more complete study of
$J/\psi$ suppression in heavy ion collisions.
In the following we will consider $\chi_c$ contributing to
about $40\%$ of the initial $J/\psi$ production and use the above 
formula to estimate its suppression in a deconfined partonic system.

In the rest frame of a deconfined parton gas, the momentum
distribution of thermal gluons will depend on the effective 
temperature $T$ with an approximate Bose-Einstein distribution,
$ f(k^0;T)\propto [\exp(k^0/T)-1]^{-1}$.
The velocity averaged dissociation cross sections for charmonia
is defined as,
\begin{equation}
\langle v_{rel}\sigma\rangle (T,\vec{p}) 
=\frac{\int d^3k v_{rel} \sigma(q^0)
  f(k^0;T)}{\int d^3kf(k^0;T)},
\end{equation}
where $v_{rel}$ is the relative velocity between $J/\psi$ and a gluon.
In this paper we are only interested in $J/\psi$ production in the
midrapidity region, so these cross sections will depend on
charmonia transverse momentum $p_T$ as well as the effective
temperature $T$.

With the velocity averaged dissociation cross sections, 
the survival probabilities of
charmonia in the 
deconfined quark-gluon plasma will have the following form,
\begin{equation}
S^{deconf.}(\vec{b},\vec{r},\vec{v})=\exp\{-\int_{\tau_0}^{\tau_f} 
\langle v_{rel}\sigma\rangle 
\rho(\vec{r}+\vec{v} \tau,\tau) \Theta (\rho-\rho_c)d \tau\},
\end{equation}
where $\tau_0$ is the formation time of the quark-gluon plasma, which will
be set as $\tau_0=1fm$ in the following calculations.
The upper limit of the time integral $\tau_f$ is determined by the
$\Theta$ function. Here we introduce the critical density $\rho_c$
above which the QGP dissociation effects enters. We assume it is the
same for both $J/\psi$ and $\chi_c$. Because of different binding
energies, the effective cross sections will have different temperature 
dependences for $J/\psi$ and $\chi_c$ even if $\rho_c$ is reached.
$\rho$ is the local density depending on $\tau$, the initial production
point $\vec{r}$ and the velocity $\vec{v}$ of the charmonium particles.
The velocity $\vec{v}$ depends on the transverse momentum $\vec{p}_T$
and the azimuthal angle $\phi$. 
The dissociation cross sections $\langle v_{rel}\sigma\rangle$ depend
on the effective temperature, which will also depend on the local
density $\rho_\tau(b,s)$.
In the case
of 1+1D Bjorken longitudinal expansion with the initial plasma density
$\rho_0=\rho(\tau=\tau_0)$,
\begin{equation}
\rho_\tau(b,s)=\rho_0(b,s) (\frac{\tau_0}{\tau})^\alpha,
\end{equation}
where $\alpha=1$.
Correspondingly, for the effective temperature,
\begin{equation}
T_\tau(b,s)=T_0(b,s) (\frac{\tau_0}{\tau})^{1/3}.
\end{equation}
For simplification, we relate the local effective temperature with the
local density in the following way,
\begin{equation}
T_\tau(b,s)=\kappa \left (\rho_\tau(b,s)\right )^{1/3},
\end{equation}
where $\kappa=[\pi^2/16\zeta(3)]^{1/3}$.

The initial plasma density is related to the rapidity density
of gluons,
\begin{equation}
\rho_0(b,s)=\frac{1}{\tau_0}\frac{dN^g}{dyd^2s}(b,s),
\end{equation}
where $d^2s$ is the transverse area of the overlapping region
of two colliding nuclei.
We will follow the two-component model \cite{wg} and include 
both the soft and hard contribution to the final hadron
production. Assuming that the initial gluon density is proportional
to the final hadron rapidity density, we have phenomenologically,
\begin{equation}
\frac{dN^g}{dyd^2s}(b,s)=c[f_b n_b(b,s)+f_p n_p(b,s)],
\end{equation}
where $c$ is a constant and we will set $c=1$ in the following
calculations, and
\begin{eqnarray}
n_p(b,s)&=&T_A(s) [1-\exp(-T_B(b-s) \sigma_{pp})]+
T_B(b-s) [1-\exp(-T_A(s) \sigma_{pp})],\\
n_b(b,s)&=&T_A(s)T_B(b-s)\sigma_{pp}.
\end{eqnarray}
Since mini-jet cross section at the SPS energy is very small, we will
effectively only have the soft contribution which is proportional
to the number of participant nucleons. At collider energies such 
as RHIC the contribution from the hard processes is more
important. We will use $f_b=0.34$ and $f_p=0.88$ as determined by the
PHENIX \cite{phinex} experiment for $Au+Au$ collisions at $\sqrt{s}=130$ 
GeV. We extrapolate these parameters to $\sqrt{s}=200GeV$ by just 
multiplying a factor of $1.14$ found by PHOBOS \cite{phobos}. 

The final expression for the survival probability due to QGP 
suppression is,
\begin{equation}
S^{deconf.}(\vec{b},\vec{r},\vec{v})=\exp\{-
\int_{\tau_0}^{\tau_f} \frac{d\tau}{\tau}
\langle v_{rel}\sigma_{\psi g}\rangle (\vec{r}+\vec{v} \tau,\tau)
\rho_0(b,\vec{r}+\vec{v} \tau)) \Theta(\rho-\rho_c)\}.
\end{equation}
It is interesting to note that with a very large constant dissociation cross
section the above formula will be equivalent to the model\cite{blaizot} by
Blaizot {\it et al.}, where they assume that all of $J/\psi$ will be 
dissociated above some critical density. The detailed comparison of 
this approach and ours will be presented elsewhere.

We can also include the transverse expansion effects on the local
parton density following Ref.~\cite{gvwh},
\begin{equation}
\rho(\vec{r},\tau)=\frac{\tau_0}{\tau}
\int\frac{d\Omega_{v_t}}{2\pi}\rho_0(\vec{r}-\vec{v}_t \tau),
\end{equation}
where $v_t$ is the average velocity of the transverse expansion of the
parton system. We will use $v_t=0.4c$ for SPS and $v_t=0.6c$ for RHIC
in the following numerical calculations.

Apart from the above discussed QGP suppression for charmonia states,
there is also suppression associated with the initial state interaction, i.e., 
the nuclear absorption of so-called preresonance of $c\bar c$ pairs,
\begin{equation}
S^{abs}(\vec{b},\vec{r})=\frac{(1-\exp(-\sigma_{abs} T_A(r)))
  (1-\exp(-\sigma_{abs} T_B(|\vec{b}-\vec{r}|)))}
  {\sigma_{abs}^2T_A(r)T_B(|\vec{b}-\vec{r}|)}.
\end{equation}
where $\sigma_{abs}$ is the absorption cross section of the
preresonance with nucleons, for which we will set
$\sigma_{abs}=5.8mb$ in this paper.

By summing up these two contributions, we get the final 
$J/\psi$ survival probability  as
\begin{equation}
S^{sur.}(\vec{b},\vec{p}_T)=\frac{\int d^2\vec{r}T_A(r)T_B(|\vec{b}-\vec{r}|)
  S^{abs}(\vec{b},\vec{r})S^{deconf.}(\vec{r},\vec{v})}
  {\int d^2\vec{r}T_A(r)T_B(|\vec{b}-\vec{r}|)}.
\end{equation}
From the above expression, we see that the nuclear absorption has no
dependence on the azimuthal angle. This means that there is no 
contribution to azimuthal anisotropy from the initial nuclear absorption. 
On the other hand, the final state
interaction or the dissociation by the deconfined parton gas 
indeed has azimuthal angular dependence because the parton
density is azimuthally asymmetric. Therefore, any finite value of
angular anisotropy for $J/\psi$ production should come from the final state 
interaction. It will provide us important information about the 
early stage of the quark-gluon plasma.

We will quantify the azimuthal anisotropy by the second Fourier 
coefficient $v_2$ of the azimuthal angle distribution of the 
final $J/\psi$ distribution, similarly to the proposed 
elliptic flow measurement \cite{oll,posk}.
It is defined as
\begin{equation}
v_2(b,p_T)=\frac{\int d\phi S^{sur.}(\vec{b},\vec{p}_T) \cos
  (2\phi)}
  {\int d\phi S^{sur.}(\vec{b},\vec{p}_T) },
\end{equation}
where $\phi$ is the azimuthal angle between $J/\psi$ transverse
momentum $\vec{p}_T$ and the impact parameter $\vec{b}$.
With this formula we can study both the $p_T$ dependence and the
centrality ($b$) dependence of $v_2$.

In the following we present the numerical results of the above 
approach. We first determine the critical density $\rho_c$ by 
fitting the SPS data on $J/\psi$
suppression, and then predict the suppression and $v_2$ at RHIC.

In Fig.~1, we show our results at the SPS energy. The upper plot is
the survival probability as a function of transverse energy $E_T$
compared with the experimental data from NA50\cite{na50}. From the fit
we determined the critical density $\rho_c=3.3fm^{-3}$.
The correlation between the impact parameter $b$ and transverse
energy $E_T$ \cite{kharzeev1},
$$P(E_T,b)=\frac{1}{\sqrt{2\pi q^2aN_p(b)}}\exp\left\{-\frac{[E_T-qN_p(b)]^2}
{2q^2aN_P(b)}\right\},
$$
has been used in the calculations,
where we set the parameters as $q=0.274GeV$ and $a=1.27$\cite{blaizot}.
In the calculations, we also include the $E_T$ fluctuation effects as 
in\cite{blaizot,capella,polleri}, which is important for the last
few $E_T$ bins. From this figure, we can see that 
our approach can well describe the
experimental data of NA50, and taking into account the transverse
expansion improves the fit especially in the peripheral region.
As expected, $v_2$ vanishes at very peripheral collisions, and becomes
sizable when the anomalous suppression makes sense. For more central
collisions, because of the symmetric geometry of the collisions, 
the azimuthal asymmetry $v_2$ vanishes again.
Since we assumed that only final state interactions produce
finite value of $v_2$, its increase with $E_T$ is quite abrupt
around the value when the critical density $\rho_c$ is reached. 
After taking into account the transverse expansion, $v_2$ is a
little higher than the case without transverse expansion. 
This is quite different from $v_2$ for jet quenching\cite{gvwh}, 
where transverse expansion is found to reduce $v_2$ significantly. 

The results at RHIC are shown in Fig.~2. The survival probability 
and $v_2$ of $J/\psi$ are shown as functions of number of participants.
Comparing with the results in Fig.~1, we find that the 
anomalous suppression enters already at peripheral collisions
at RHIC, and the gap between the full
suppression and the nuclear absorption alone is much larger than that 
at SPS. And $v_2$ is also much larger.

It is interesting to note the different dependences of $v_2$ on the
the transverse expansion at the SPS and RHIC energies. 
At SPS, it enhances $v_2$ a little bit while at RHIC it reduces. 
This is because at these two different energies the dominant suppression 
sources are different in the region where $v_2$ is sizable.
At SPS, a large part of $J/\psi$ suppression comes 
from $\chi_c$ suppression, because the density is not so high and 
it is difficult to dissociate directly produced $J/\psi$. 
For this part, the dissociation cross section is very large,
in the order of a few $mb$. So $\chi_c$ would be dissociated
almost totally  when the local density is above the critical
value $\rho_c$, which is more similar to the Blaizot-like model.
In this case, the transverse expansion will increase $v_2$ a 
little bit because it increases the time duration when $\chi_c$ is
being dissociated. This increase of time duration even overcomes the 
decrease of geometrical asymmetry due to transverse expansion
leading to an increased $v_2$. The dominant suppression, however,
comes from directly produced $J/\psi$ (about $60\%$) at the RHIC 
energies because of the much higher initial density. 
The transverse expansion accelerates the decrease of the initial
density and reduces the initial geometric asymmetry. This
leads to the decrease of $v_2$, very similar to the case of jet 
quenching in dense. Since the final total $J/\psi$ suppression
is a mixture of direct suppression and suppression through $\chi_c$, 
this reduction is not as large as it is for jet quenching \cite{gvwh}.

In conclusion, we have studied the azimuthal asymmetry of $J/\psi$ 
suppression at both SPS and RHIC energies within a dynamic model
of charmonia dissociation in a deconfined partonic system.
We assume the hadronic absorption in our model is negligible so
that the only final state $J/\psi$ absorption is due to
gluon dissociation in deconfined medium above a critical
parton density.
With a critical density $\rho_c=3.3fm^{-3}$ we can reproduce well 
the anomalous suppression found by the NA50 experiment at the SPS.
We predicted the azimuthal anisotropy $v_2$ of the $J/\psi$ suppression
at SPS. The existence of a critical density $\rho_c$ for $J/\psi$
suppression leads to a sharp increase of $v_2$ with $E_T$, assuming
that $v_2$ only comes from $J/\psi$ dissociation in the deconfined 
matter. At the RHIC energies, we found that the anomalous suppression 
already plays a role in peripheral collisions because of high density. 
In noncentral collisions there is sizable $v_2$ for $J/\psi$ at $p_T=3GeV$.

The experimental study of $v_2$ for $J/\psi$ suppression will be 
complimentary to other studies of $J/\psi$ suppression.
Together with measurements, such as high $p_T$ hadrons where jet
quenching plays an important role\cite{v2-pt}, these studies will 
provide valuable information about the early stage of high-energy
heavy-ion collisions.

\noindent {\bf Acknowledgments}: We thank J.~H\"ufner
for interesting discussions and critical reading of the manuscript.
This work was supported by 
the Director, Office of Energy Research, Office of High Energy 
and Nuclear Physics, Divisions of Nuclear Physics, of the U.S. 
Department of Energy under Contract No. DE-AC03-76SF00098 
and in part by NSFC under project No. 19928511.

\begin{figure}[t]
\centerline{\psfig{figure=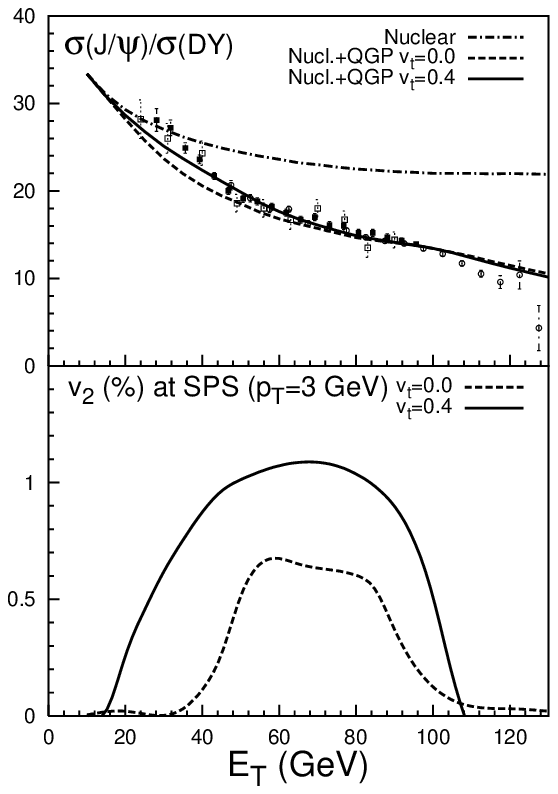,width=10cm,height=16cm}}
\vskip 0.2cm
\caption{The $J/\psi$ survival probability and $v_2$ at SPS as a function of
transverse energy $E_T$: nuclear absorption alone (dot-dashed
line); nuclear absorption plus QGP dissociation without (dashed line) 
and with transverse expansion effects. 
The experimental data are from NA50\protect\cite{na50}. }
\end{figure}

\vskip 0.2cm

\begin{figure}[t]
\centerline{\psfig{figure=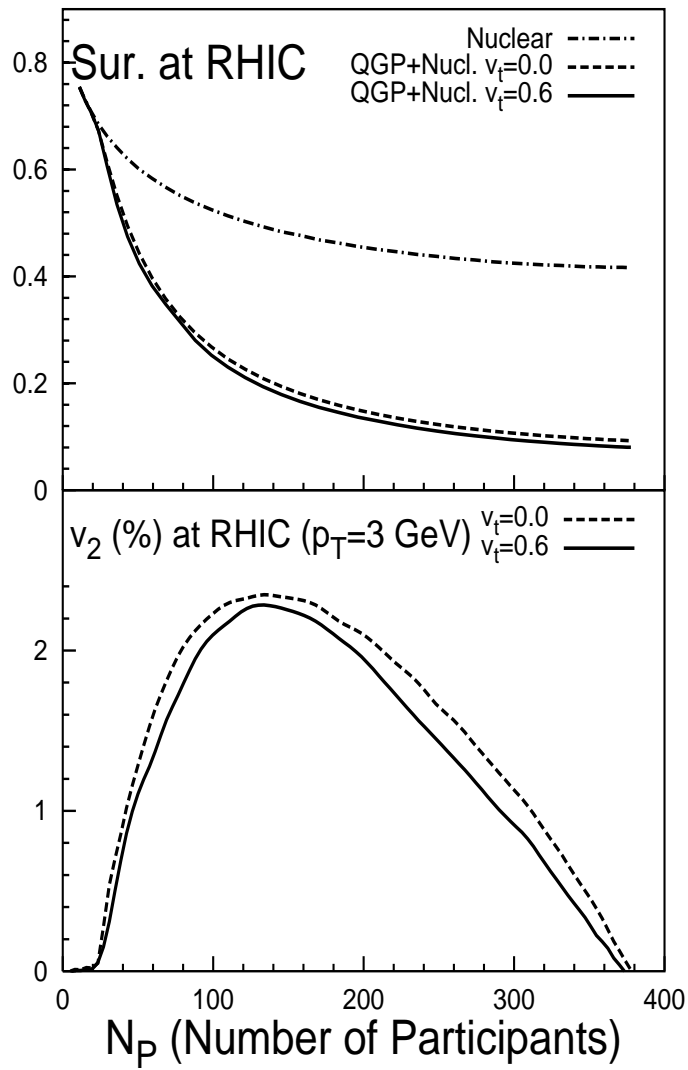,width=10cm,height=16cm}}
\vskip 0.2cm
\caption{The $J/\psi$ survival probability and $v_2$ at RHIC as a function of
number of participants $N_P$.}
\end{figure}

\end{document}